\documentclass[aps,prb,a4paper,superscriptaddress,twocolumn,floatfix]{revtex4}
\usepackage{color}
\usepackage{graphicx}
\usepackage{latexsym}
\usepackage[]{amsmath}
\usepackage{epstopdf}

\begin{document}
\newcommand{\chem}[1]{\ensuremath{\mathrm{#1}}}

\title{Measurement of the internal magnetic field in the correlated iridates Ca$_4$IrO$_6$, Ca$_5$Ir$_3$O$_{12}$, Sr$_3$Ir$_2$O$_7$ and Sr$_2$IrO$_4$}

\author{I.\ Franke}
\affiliation{Oxford University Department of Physics, Clarendon Laboratory,
Parks Road, Oxford OX1 3PU, United Kingdom}

\author{P.\ J.\ Baker}
\affiliation{Oxford University Department of Physics, Clarendon Laboratory,
Parks Road, Oxford OX1 3PU, United Kingdom}
\affiliation{ISIS Pulsed Neutron and Muon Source, STFC Rutherford Appleton Laboratory, 
Harwell Science and Innovation Campus, Didcot, Oxfordshire, OX11 0QX, United Kingdom}

\author{S.\ J.\ Blundell}
\affiliation{Oxford University Department of Physics, Clarendon Laboratory,
Parks Road, Oxford OX1 3PU, United Kingdom}

\author{T.\ Lancaster}
\affiliation{Oxford University Department of Physics, Clarendon Laboratory,
Parks Road, Oxford OX1 3PU, United Kingdom}

\author{W.\ Hayes}
\affiliation{Oxford University Department of Physics, Clarendon Laboratory,
Parks Road, Oxford OX1 3PU, United Kingdom}

\author{F.\ L.\ Pratt}
\affiliation{ISIS Pulsed Neutron and Muon Source, STFC Rutherford Appleton Laboratory, 
Harwell Science and Innovation Campus, Didcot, Oxfordshire, OX11 0QX, United Kingdom}

\author{G.\ Cao}
\affiliation{Department of Physics and Astronomy, University of Kentucky, Lexington, KY 40506}

\date{\today}

\begin{abstract}
Oxides containing iridium ions display a range of magnetic and
conducting properties that depend on the delicate balance between
interactions and are controlled, at least in part, by the details of
the crystal architecture.  We have used muon-spin rotation ($\mu$SR)
to study the local field in four iridium oxides, Ca$_4$IrO$_6$,
Ca$_5$Ir$_3$O$_{12}$, Sr$_3$Ir$_2$O$_7$ and Sr$_2$IrO$_4$, which show
contrasting behavior.  Our $\mu$SR data on Ca$_4$IrO$_6$ and
Ca$_5$Ir$_3$O$_{12}$ are consistent with conventional
antiferromagnetism where quasistatic magnetic order develops below
$T_{\rm N}=13.85(6)$~K and 7.84(7)~K respectively.  A lower internal
field is observed for Ca$_5$Ir$_3$O$_{12}$, as compared to
Ca$_4$IrO$_6$ reflecting the presence of both Ir$^{4+}$
and Ir$^{5+}$ ions, resulting in a more
magnetically dilute structure.  Muon precession is only observed over
a restricted range of temperature in Sr$_3$Ir$_2$O$_7$, while the Mott
insulator Sr$_2$IrO$_4$ displays more complex behavior, with the
$\mu$SR signal containing a single, well-resolved precession signal
below $T_{\rm N}=230$\,K, which splits into two precession signals at
low temperature following a reorientation of the spins in the ordered
state.
\end{abstract}

\maketitle

\section{Introduction}
Transition metal oxides have attracted widespread interest in recent
years, driven partly by the observation of high temperature
superconductivity in cuprates, colossal magnetoresistance in
manganites and p-wave superconductvity in
ruthenates\cite{dagotto_01,imada_01}. A wide variety of behavior can
be expected since the transition metal ion can be any one of the 3d,
4d or 5d series and the possible crystal architectures that can be
synthesized include chains, ladders, square layers, triangular layers,
kagome layers, pyrochlore lattices and many more. The 5d series has
received the least attention but the extended nature of the 5d
orbitals leads to large crystal field energies\cite{cao_04}. The
spin-orbit interaction is also strong, owing to the large atomic
number, and these competing energy scales are likely to open up new
possibilities for their magnetic behavior.

The iridium ion is a particularly attractive candidate for study
because Ir$^{4+}$ (5d$^5$) is an $S=\frac{1}{2}$ species and iridates
display a wide variety of unusual characteristics.  For example,
\chem{Ca_{4}IrO_{6}} and \chem{Ca_{5}Ir_{3}O_{12}} are
antiferromagnetically ordered at low temperature but exhibit weak
ferromagnetism due to spin canting\cite{cao_04, wakeshima_01}.
\chem{Ca_{5}Ir_{3}O_{12}} exhibits a more complex partial spin
ordering due to the presence of both Ir$^{4+}$ and Ir$^{5+}$
ions\cite{cao_04}.  These two compounds have a large value of the
ratio $\theta/T_{\rm N}$ where $\theta$ is the Curie-Weiss temperature
(see Table~\ref{tab:values}) demonstrating the presence of competing
interactions.  \chem{Sr_{3}Ir_{2}O_{7}} is a weak ferromagnet showing
complicated crossovers in its magnetic behaviour that have not yet
been explained at the microscopic level\cite{cao_06}.
\chem{Sr_{2}IrO_{4}} is a Mott insulator driven from the metallic
state by the spin-orbit coupling to a $J_{\rm eff}=\frac{1}{2}$
state\cite{kim_01,kim_02}. It is also a weak ferromagnet, shows a
novel magnetoelectric state with a so far unexplained magnetic
transition\cite{Chikara, korneta} and has even been suggested as a
candidate for high-temperature superconductivity if doped
\cite{wangsenthil}.
 
In order to study systematically the dependence of local magnetic
properties on crystal architecture we have used muon spin rotation
($\mu$SR) to investigate the microscopic magnetism and the critical
behavior of each of these compounds.  The nature of the muon spin
rotation technique makes it very sensitive to the local magnetic
environment, and especially to short-range magnetic order.  We find a
comparatively conventional development of the internal magnetic field
in \chem{Ca_{4}IrO_{6}} with some more complex features in
\chem{Ca_{5}Ir_{3}O_{12}}.  Compared to these the results for
\chem{Sr_{3}Ir_{2}O_{7}} and \chem{Sr_{2}IrO_{4}} reveal very
different behavior.  Two magnetic transitions are evident in our data
for \chem{Sr_{3}Ir_{2}O_{7}}, while in the Mott insulator
\chem{Sr_{2}IrO_{4}} we observe the development of spin reorientation
at low temperature.

\section{Experimental methods}
Small crystallites of the materials were grown in Pt crucibles using
self-flux techniques from off-stoichiometric quantities of
\chem{IrO_2}, \chem{SrCO_3} or \chem{CaCO_3}, and \chem{SrCl_2} or
\chem{CaCl_2}. These mixtures were heated to 1480$^\circ$C in Pt
crucibles, fired for 20 h and slowly cooled at 3$^\circ$C/hour to
lower temperatures (1440$^\circ$C for \chem{Sr_3Ir_2O_7}, for
example). Crystal structures of all single crystals studied were
determined at 90~K and 295~K using a Nonius Kappa CCD X-Ray
Diffractometer and Mo K(alpha) radiation. Chemical compositions of the
single crystals were determined using energy dispersive X-ray analysis
(EDX)
\begin{table*}[t]
\centering
\begin{tabular}{lcrlrlcccccc}
\hline \hline &space group & $\mu_{\rm eff}$ & & $\theta$ & & $T_{\rm
  N }$ & $\vert\theta/T_{\rm N}\vert$ & $\alpha$ & $\beta$ &
$\nu_1(0)$ & $\nu_2(0)$ \\ 
& & ($\mu_{\rm B}$/f.u) & &(K) & & (K) & &
& & (MHz) & (MHz) \\ \hline 
\chem{Ca_{4}IrO_{6}} & R$\bar{3}$c & 1.76
& [\onlinecite{sarkozy}] & $-54$ & [\onlinecite{segal_01}] &13.95(6) &
3.8(2) & 4.2(3) & 0.37(1)& 8.29(3) & 4.39(4) \\ 
\chem{Ca_{5}Ir_{3}O_{12}} & P$\bar{6}$2m& 1.5 &
   [\onlinecite{wakeshima_01}] & $-280$ & [\onlinecite{cao_04}] &
   7.84(7) & 35.7(1) & 2.8(5) &0.40(6) & 1.30(2) & --\\ 
\chem{Sr_{3}Ir_{2}O_{7}} &I4/mmm & 0.69 &
   [\onlinecite{cao_06}] &$-17$ & [\onlinecite{cao_06}] & 285 & $0.06
   $ & -- & -- & -- & -- \\ \chem{Sr_{2}IrO_{4}} &I4$_1$/acd & 0.33 &
   [\onlinecite{kini}] & 251 & [\onlinecite{kini}] & 230.4(3) &
   1.08(2) & 0.98(4) &0.21(1) & 2.93(1) & -- 
\\ \hline \hline
\end{tabular}
\caption{Properties of the four compounds studied and parameters
  extracted from the fits presented here. The values of $T_{\rm N}$
  are from this work, except for Sr$_3$Ir$_2$O$_7$, which is taken
  from Ref.~\onlinecite{cao_06}.}
\label{tab:values}
\end{table*}

Zero-field $\mu$SR measurements\cite{blundell_cont} were made using
the General Purpose Surface (GPS) muon spectrometer at the Swiss Muon
Source. Each sample was wrapped in 25~$\mu$m silver foil and mounted
on a silver backing plate. Spin-polarized positive muons are implanted
in the target sample, where the muon usually occupies a position of
high electronegativity in the crystal. The observed property in the
experiment is the time-evolution of the muon-spin polarization, which
is proportional to the positron asymmetry function $A(t)$ and depends
on the local magnetic field $B$ at the muon stopping
site\cite{blundell_cont}. In these polycrystalline samples the
internal magnetic field will be randomly orientated with respect to
the initial muon polarization. Because of this we can expect that in a
fully ordered magnet $\frac{2}{3}$ of the signal will be due to
oscillations about the magnetic field perpendicular to the muon spin
and $\frac{1}{3}$ will be relaxation due to the dynamic magnetic fluctuations
parallel to the muon spin direction\cite{blundell_cont}.

\begin{figure}[b]
\includegraphics[width=\columnwidth]{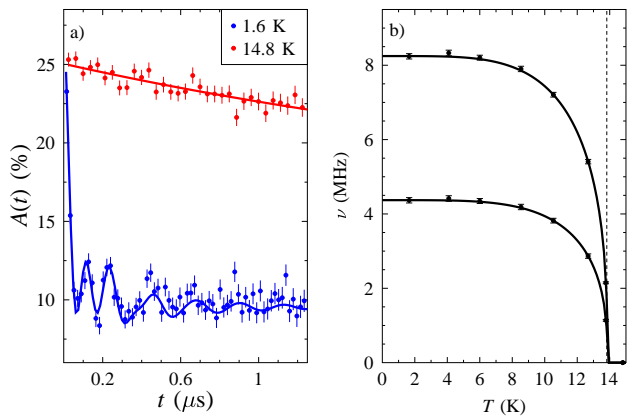}
\caption{ (Color online) (a) Raw data for \chem{Ca_{4}IrO_{6}} above
  and below the magnetic phase transition. The low temperature data
  are fitted with two oscillating components as described by
  Eq.~(\ref{eq:416}).  (b) Precession frequencies $\nu_1$ and $\nu_2$
  for the ordered state.
\label{416tot}}
\end{figure}

\section{ \chem{Ca_{4}IrO_{6}} and \chem{Ca_{5}Ir_{3}O_{12}}}
\chem{Ca_{4}IrO_{6}} has a rhombohedral structure, in which 1D chains
of \chem{IrO_6} octahedra alternate with \chem{CaO_6} trigonal
prisms\cite{segal_01,cao_04}. In contrast \chem{Ca_{5}Ir_{3}O_{12}}
has a triangular lattice and a hexagonal
structure\cite{dijksma_01,wakeshima_01} and, unusually amongst
iridates, in addition to \chem{Ir^{4+}} it also
includes \chem{Ir^{5+}} ions.  In a cubic crystal field Ir$^{5+}$ is
in a $S=0$ state \cite{hayashi_01,darriet_01}, but in a rhombohedral
or hexagonal environment the sizeable spin-orbit coupling may modify
the level scheme.
In spite of these differences both structures lead to the formation of
spin chains perpendicular to a triangular lattice, promoting the
occurrence of frustration.

Both compounds are insulators (below 300K for
\chem{Ca_{5}Ir_{3}O_{12}}) and exhibit antiferromagnetic order. The
transition temperature for \chem{Ca_{4}IrO_{6}} was
reported\cite{cao_04,segal_01} to be in the range of 12--16~K.
\chem{Ca_{5}Ir_{3}O_{12}} is found to order into an antiferromagnetic
state below 7.8~K \cite{wakeshima_01,cao_04}.

Example data for \chem{Ca_{4}IrO_{6}} above and below the transition
are shown in Fig.~\ref{416tot}(a).  In the paramagnetic phase the
signal is described by a single exponential relaxation and below
$T_{\rm N}$ we observe two well-defined precession frequencies along
with a fast relaxing component at early times.  The data in this
regime may be parametrized using the equation:
\begin{eqnarray}
A(t) &=& A_1 \exp(-\lambda_{1}t)\cos(2\pi\nu_1 t) \nonumber \\ 
&+& A_2 \exp(-\lambda_{2}t)\cos(2\pi\nu_2 t) \nonumber \\
&+& A_3 \exp(-\Lambda t), + A_{\rm bg} 
\label{eq:416}
\end{eqnarray}
where $A_i$ are the amplitudes of the components, $\lambda_{ i}$ are
the corresponding relaxation rates, $\nu_i$ are the precession
frequencies, and $\Lambda$ is the relaxation rate of the fast-relaxing
component.  The constant $A_{\rm bg}$ accounts for muons stopped
outside the sample as well as the expected one-third contributions
arising from muon-spin components directed parallel to the local
magnetic field.  A satisfactory fit to the data could be obtained by
fixing $A_1$, $A_2$, $A_3$, $\lambda_1$ and $\lambda_2$ to
temperature-independent values (4.02\%, 1.37\%, 11.2\%, 5.03~MHz and
1.55~MHz respectively) and allowing $\nu_1$ and $\nu_2$ to vary in a
fixed proportion $\nu_2/\nu_1=0.53$ (a ratio determined to better than
4\% accuracy by first allowing them to vary independently).  In
addition, the large relaxation rate $\Lambda$ was found to scale as
$\nu_1^2$ and reaches 22\,MHz at 1.5\,K.  We deduce that there are two
independent muon sites which experience a quasistatic local field
(leading to the oscillatory components) and a further site at which
the muon spin relaxes in a manner dominated by the magnitude of the
local field (leading to the fast-relaxing component).  The
fast-relaxing component observed below the magnetic ordering
temperature probably reflects a class of muon site at which the local
field is large but where there exists some degree of spatial or
temporal disorder. The results of fitting the frequency $\nu_1$ are
illustrated in Fig.~\ref{416tot} (b).

The  temperature dependence of $\nu_1$ was fitted to the phenomenological form 
\begin{equation}
\nu_1(T) = \nu_1(0)[1-(T/T_{\rm N})^{\alpha}]^{\beta}.
\label{eq:alphabeta}
\end{equation} 
The parameters extracted were: $T_{\rm N} = 13.85(6)$~K, $\nu_1(0)
=8.29(3)$~MHz, $\alpha = 4.2(3)$, and $\beta = 0.37(1)$,  and this
value of $\beta$ is consistent with a typical three-dimensional
magnetic order parameter\cite{blundell}, though in this case there are
insufficient data close to $T_{\rm N}$ to provide an unambiguous
determination of $\beta$.

\begin{figure}[t]
\includegraphics[width=\columnwidth]{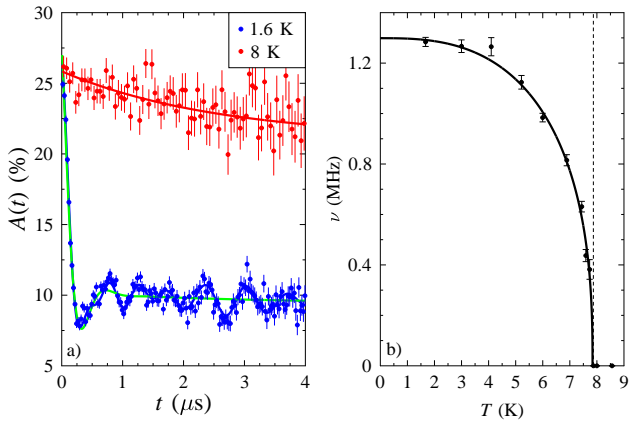}
\caption{ 
(Color online) 
(a) Raw data for \chem{Ca_{5}Ir_{3}O_{12}} above and below the magnetic phase transition. The data below the transition is fitted with a single frequency (green) and a five frequency (blue) fit. 
(b) Precession frequency extracted from a single frequency fit to the \chem{Ca_{5}Ir_{3}O_{12}} data.
\label{5312tot}}
\end{figure}

The form of the raw data for \chem{Ca_{5}Ir_{3}O_{12}} is more
complicated. Though dominated by a single primary oscillation
frequency, fitting with additional frequency components with smaller
amplitudes (and frequencies ranging from 0.8--4 MHz) improved the
quality of the fit, using up to four or five
frequencies. Fig.~\ref{5312tot} shows the data above and below the
transition, together with a fit to the low temperature data using one
and five frequencies for comparison. While the five frequency fit
clearly succeeds in describing the data over a significantly longer
time, the parameters of the smaller amplitude oscillations are not
well defined and the frequencies could not be followed through to
higher temperatures. The single frequency fit achieves a satisfactory
and consistent parametrization of the largest amplitude oscillating
component across the temperature range measured.  These results
indicate the existence of several magnetically inequivalent muon
stopping sites.

The data set was analysed by fitting just to the frequency component
with the largest amplitude, the precession frequency of which is shown
in Fig.~\ref{5312tot} and reaches $\approx 1.3$~MHz as $T \rightarrow
0$.  This precession frequency was fitted to Eq.~(\ref{eq:alphabeta}),
yielding $T_{\rm N} = 7.84(7)$~K, $\nu(0) = 1.30(2)$~MHz, $\alpha =
2.8(5)$ and $\beta = 0.40(6)$, and these parameters are also
consistent with three-dimensional behavior.

\section{ \chem{Sr_{2}IrO_{4}} and \chem{Sr_{3}Ir_{2}O_{7}}}

\begin{figure}[t]
\includegraphics[width=\columnwidth]{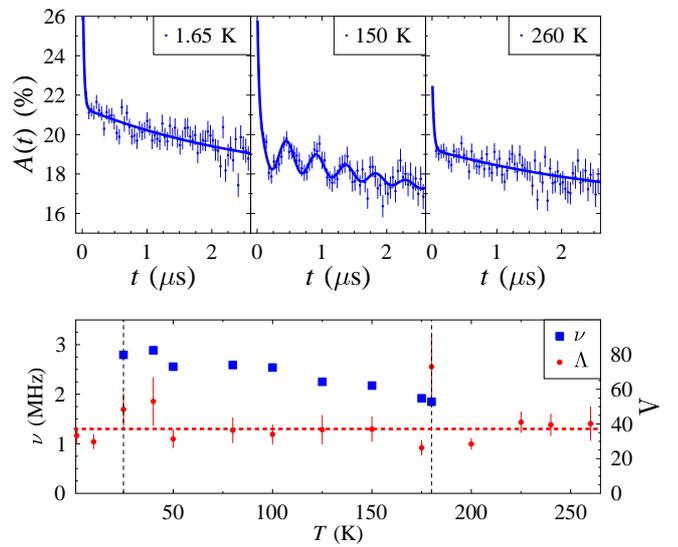}
\caption{ 
(Color online) 
Data for \chem{Sr_{3}Ir_{2}O_{7}} in the three different phases of the material. The data  in the intermediate phase was fitted with an oscillating component $\nu$ (blue squares) and a relaxing component $\Lambda$ (red circles) [Eq.~(\ref{eq:5312})]. The relaxation $\Lambda$ (red circles) can be observed over the whole range of temperatures.
\label{327panel}}
\end{figure}

\chem{Sr_{3}Ir_{2}O_{7}} and \chem{Sr_{2}IrO_{4}} are both
antiferromagnetic with transition temperatures $T_{\rm N}$ at 285~K
(Ref.~\onlinecite{cao_06}) and 240~K (Ref.~\onlinecite{cao_01}),
respectively.  Both compounds contain Ir$^{4+}$ ($S=\frac{1}{2}$)
ions.  In both compounds, a weak ferromagnetic moment is observed
suggesting that they are both canted
antiferromagnets\cite{kim_01,kini,nagai}.

\chem{Sr_{3}Ir_{2}O_{7}} belongs to the Ruddlesden-Popper series
\chem{Sr_{n+1}Ir_{n}O_{3n+1}} with $n=2$ and is constructed out of
Ir-O bilayers with Sr-O interlayers\cite{cao_06}.  Field cooled
magnetization data picked up three anomalies, a kink at $T^{\rm *}$ =
260~K, a steep downturn at $T_{\rm D}$ = 50~K and the magnetization
becoming negative\cite{cao_06} below about 20~K.  These features were
not observed in the zero-field-cooled magnetization data, implying a
strong spin disordering or a random orientation of magnetic domains
that persists through $T_{\rm N}$. The transitions at $T^{\rm *}$ and
$T_{\rm D}$ could also be identified in resistivity data, but no
explanation for the comprehensive magnetization and transport behavior
has been found so far\cite{cao_06}.

\begin{figure}[t]
\includegraphics[width=\columnwidth]{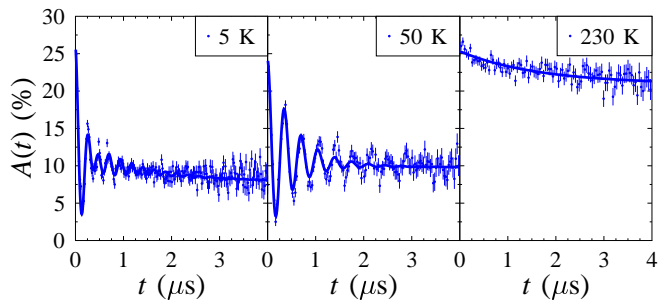}
\caption{ (Color online) Raw data for \chem{Sr_{2}IrO_{4}} above and
  below the magnetic phase transition. The low temperature data was
  fitted with one relaxation component and two oscillating components
  [Eq.~(\ref{eq:416})], whereas for the data above 20~K we used a
  relaxing component and only one oscillating component
  [Eq.~(\ref{eq:5312})].
\label{214panel}}
\end{figure}

\chem{Sr_{2}IrO_{4}} is the $n=1$ member of the Ruddlesden-Popper
series (i.e.\ it has the \chem{K_{2}NiF_{4}} structure) and is
therefore constructed out of single Ir-O layers with Sr-O interlayers
\cite{crawford}.  The strong spin-orbit coupling acting on the
$t_{2g}$ manifold results in a filled $J=\frac{3}{2}$ lower band and a
half-filled $J=\frac{1}{2}$ upper band.  The narrow bandwidth of the
upper band leads to a Mott insulating state, even though the
electronic correlations (parametrized by $U$) are weaker than in more
familiar Mott insulators, such as the underdoped cuprates.
Sr$_2$IrO$_4$ can therefore be thought of as a spin-orbit induced Mott
insulator \cite{kim_01,moon_01,jin,watanabe}.  In addition, a
metamagnetic transition occurs at 0.2~T, well below
240~K\cite{kim_02}. Chikara {\em et al.}\cite{Chikara} have recently
discovered another magnetic transition at 100 K, which is thought to
be a reorientation transition of the spins.

The results of our $\mu$SR experiments on \chem{Sr_{3}Ir_{2}O_{7}} are
summarized in Fig.~\ref{327panel} where the muon decay asymmetry for
1.5~K, 150~K and 260~K is presented. The presence of a fast relaxing
component observed at short times can be taken to be a signature of
magnetic order, but oscillations with a single precession frequency
are only observed between 20~K and 160~K. In this temperature interval
we fitted the data using
\begin{equation}
A(t) = A_1 \exp(-\lambda_1t)\cos(2\pi\nu_1 t) + A_2 \exp(-\Lambda t).
\label{eq:5312}
\end{equation}
The disappearance of the precession signal below 20~K is unusual and
corresponds to the temperature at which the magnetization was observed
to become negative\cite{cao_06}. We note that it was not possible to
fit the frequency to Eq.~(\ref{eq:alphabeta}) since the oscillations
are only observed outside the critical region.

\begin{figure}[t]
\includegraphics[width=\columnwidth, height=9cm]{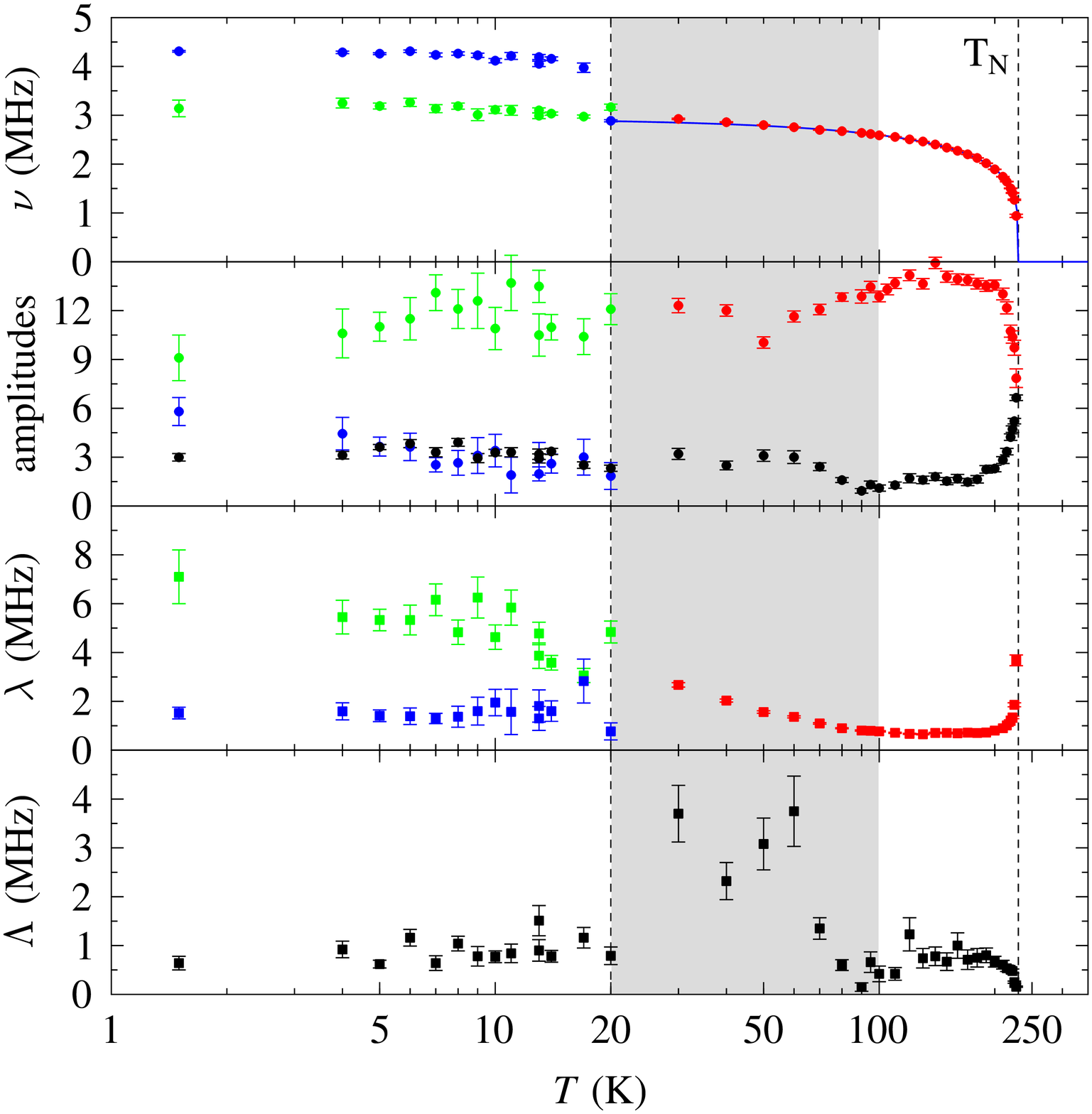}
\caption{ (Color online) Fitted frequencies, amplitudes, linewidths
  and relaxation of \chem{Sr_{2}IrO_{4}}. The amplitudes and
  linewidths corresponding to the different components have been
  differentiated by color.  The shaded region marks the temperature
  interval in which the relaxation rates increase on cooling, before
  the single precession frequency splits into two.
\label{214log}}
\end{figure}

\begin{figure}[t]
\includegraphics[width=\columnwidth]{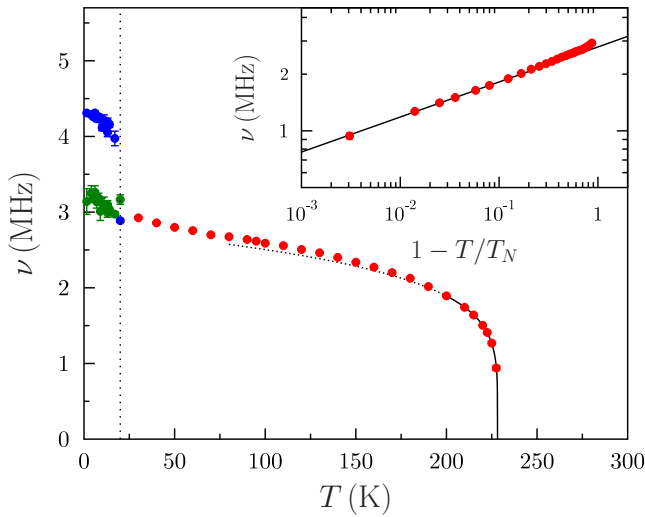}
\caption{ 
(Color online) 
Temperature dependence of the fitted frequency observed in
\chem{Sr_{2}IrO_{4}}. Inset: a fit of these data near $T_{\rm N}$.
\label{214new}}
\end{figure}

The data obtained for \chem{Sr_{2}IrO_{4}} show much richer
behavior. The $\mu$SR spectra at three different temperatures are
shown in Fig.~\ref{214panel}. Above the magnetic ordering transition
there is no oscillatory signal and, as is generally the case in
paramagnets, the data are well described by an exponential
relaxation. Between $T_{\rm N}$ and 20~K the data could be
parametrized using one oscillation component and one exponentially
relaxing component. Below 100~K the relaxation rates $\Lambda$ and
$\lambda_1$ both increase, the former quite sharply, and below 20~K a
second frequency needs to be included for a satisfactory fit.
The values obtained with this fitting procedure are plotted in
Fig.~\ref{214log}. The higher of the two frequencies has a smaller
amplitude and also has the smaller linewidth while the lower frequency
is rather similar in magnitude to that observed in
\chem{Sr_{3}Ir_{2}O_{7}}, suggesting a rather similar muon site within
the perovskite block in both compounds.  As the temperature decreases
in Sr$_2$IrO$_4$ the precession signal appears to evolve continuously
in frequency, amplitude and relaxation rate, transforming into the
lower frequency component seen in Fig.~\ref{214log}.  The new
component which appears below 20\,K represents additional amplitude.
The temperature interval in which the relaxation rates are growing
upon cooling, just before the second frequency appears, is shown as a
shaded region in Fig.~\ref{214log}.  The single precession frequency
was fitted with Eq.~(\ref{eq:alphabeta}) in the temperature region
from 20~K up to $T_{\rm N}$ and yields $T_{\rm N} = 230.4(3)$~K,
$\nu(0) = 2.93(1)$~MHz, $\alpha = 0.98(4)$ and $\beta = 0.21(1)$. 
Fitting just the data points close to $T_{\rm N}$ to the form $\nu(T)\propto
(1-T/T_{\rm N})^\beta$ yields $T_{\rm N}=228.2(1)$~K and
$\beta=0.19(1)$ (see Fig.~\ref{214new} and inset).
Both these
estimates of $\beta$ are consistent with a two-dimensional magnetic order
parameter\cite{blundell}.

\section{Discussion}
The results of these experiments confirm the occurrence of
antiferromagnetic order in \chem{Ca_{4}IrO_{6}} and
\chem{Ca_{5}Ir_{3}O_{12}} and provide very accurate measurements of
the N\'eel temperature in each case.  The more complex nature of the
muon signal in \chem{Ca_{5}Ir_{3}O_{12}} as compared with that in
\chem{Ca_{4}IrO_{6}} can be attributed to the larger number of
magnetically inequivalent muon stopping sites.  The presence of both
Ir$^{4+}$ and Ir$^{5+}$ in \chem{Ca_{5}Ir_{3}O_{12}} means that the
electronic moments are more dilute and this is consistent with the
smaller magnitude of the internal field measured by the muons, almost
half that found for \chem{Ca_{4}IrO_{6}}.

\chem{Sr_{3}Ir_{2}O_{7}} and \chem{Sr_{2}IrO_{4}} exhibit much more
surprising and complicated behavior.  In \chem{Sr_{3}Ir_{2}O_{7}} a
single frequency was observed between 20 and 160~K.  The disappearance
of precession below 20~K and above 160~K is mysterious.  The change
below 20~K does not correlate with any previously identified magnetic
transition but corresponds to the temperature at which the
field-cooled magnetization becomes negative.

Although the high temperature behavior of \chem{Sr_{2}IrO_{4}} is
conventional, with the appearance of one frequency below the
transition temperature at 260~K, the low temperature behavior is very
unusual.  Cooling below around 100~K induces a marked change in
behavior, resulting in the development of a second precession signal
which fully establishes below 20~K.  It has previously been
suggested\cite{Chikara} that a reorientation transition occurs at
100~K in this material, and this is compatible with our data and
correlates with a change in the Ir--O--Ir bond-angle.  In this
picture, the magnetic structure changes due to a temperature-induced
change in couplings, and this causes two structurally equivalent muon
sites to experience increasingly distinct local fields which lock in
below 20\,K.  This development of a reoriented phase may be gradual
(occurring over the shaded region of Fig.~\ref{214log}), which could
result from the balance of competing energies changing with
temperature.  This reorientation of the spins to a lower symmetry
state could be at the root of the magnetoelectric behavior
\cite{Chikara}.

\acknowledgments
We thank EPSRC for financial support and Peter Battle for useful
discussions.  
Part of this work was carried
out at the Swiss Muon Source (S$\mu$S) and we are grateful to Alex
Amato for technical support and Jack Wright for experimental assistance. GC was supported by NSF through grants DMR-0856234 and EPS-0814194.

\end{document}